\documentclass[12pt]{article}
\usepackage{amsfonts}
\usepackage{amsmath}
\usepackage{amssymb}
\usepackage{graphicx}
\usepackage{color}
\usepackage[all, knot]{xy}
\usepackage{tikz}
\usepackage{array}

\usepackage{ulem}

\usepackage[utf8]{inputenc}
\usepackage{epstopdf}
\usepackage[footnotesize]{caption}
\usepackage{amsthm}
\usepackage{enumitem}
\usepackage{mathrsfs}

\usepackage[margin=3cm]{geometry}

\def \be {\begin{equation}}
\def \ee {\end{equation}}
\def \bea {\begin{eqnarray}}
\def \eea {\end{eqnarray}}
\def \nn {\nonumber}

\def \rr {\raise.35ex\hbox{\small $\prime$}\kern-.17em{\mbox{\large $\imath$}}}

\def \dels {\partial\kern-.6em /\kern.1em}
\def \As {{A\kern-.5em / \kern.5em}}
\def \Ds {D\kern-.7em / \kern.5em}

\def \ks {k\kern-.5em /}
\def \ls {l\kern-.5em /}

\newcommand{\ci}[1]{}

\newcommand{\ba}{\begin{eqnarray}}
\newcommand{\ea}{\end{eqnarray}}
\newcommand{\bal}{\begin{align}}
\newcommand{\eal}{\end{align}}
\newcommand{\bay}[1]{\left(\begin{array}{#1}}
\newcommand{\eay}{\end{array}\right)}

\newcommand{\hide}[1]{}

\newlist{axioms}{enumerate}{2}
\setlist[axioms,1]{label=\textbf{A\arabic{axiomsi}.}, ref=A\arabic{axiomsi}}
\setlist[axioms,2]{label=\textbf{A\arabic{axiomsi}\rlap{\myEnumCounter{axiomsii}}.},%
                   ref=A\arabic{axiomsi}\myEnumCounter{axiomsii},%
                   align=parleft,%
                   leftmargin=0em,%
                   itemsep=1.4ex,%
                   before={\stepcounter{axiomsi}}}

  \usetikzlibrary{decorations.markings}

\begin{document}
\begin{titlepage}

\begin{center}

\textbf{\LARGE
Tripartite Entanglement and\\
Quantum Correlation
\vskip.3cm
}
\vskip .5in
{\large
Xingyu Guo$^{a, b}$ \footnote{e-mail address: guoxy@m.scnu.edu.cn}
and Chen-Te Ma$^{a, b, c, d}$ \footnote{e-mail address: yefgst@gmail.com}
\\
\vskip 1mm
}
{\sl
$^a$
Guangdong Provincial Key Laboratory of Nuclear Science,\\
 Institute of Quantum Matter,
South China Normal University, Guangzhou 510006, Guangdong, China.
\\
$^b$
Guangdong-Hong Kong Joint Laboratory of Quantum Matter,\\
 Southern Nuclear Science Computing Center, 
South China Normal University, Guangzhou 510006, Guangdong, China.
\\
$^c$
School of Physics and Telecommunication Engineering,\\ 
South China Normal University, Guangzhou 510006, Guangdong, China.
\\
$^d$
The Laboratory for Quantum Gravity and Strings,\\
 Department of Mathematics and Applied Mathematics,\\
University of Cape Town, Private Bag, Rondebosch 7700, South Africa.
}
\\
\vskip 1mm
\vspace{40pt}
\end{center}
\begin{abstract}
We provide an analytical tripartite-study from the generalized $R$-matrix. 
It provides the upper bound of the maximum violation of Mermin's inequality. 
For a generic 2-qubit pure state, the concurrence or $R$-matrix characterizes the maximum violation of Bell's inequality. 
Therefore, people expect that the maximum violation should be proper to quantify Quantum Entanglement. 
The $R$-matrix gives the maximum violation of Bell's inequality.
For a general 3-qubit state, we have five invariant entanglement quantities up to local unitary transformations. 
We show that the five invariant quantities describe the correlation in the generalized $R$-matrix. 
The violation of Mermin's inequality is not a proper diagnosis due to the non-monotonic behavior.
We then classify 3-qubit quantum states. 
Each classification quantifies Quantum Entanglement by the total concurrence. 
In the end, we relate the experiment correlators to Quantum Entanglement.  
\end{abstract}
\end{titlepage}

\section{Introduction}
\label{sec:1}
\noindent 
The locality on the hidden variables implies the {\it Bell's inequality} to the correlations of two separated particles \cite{Bell:1964kc}. 
Later, measuring entangled particles showed the {\it violation} of Bell's inequality \cite{Clauser:1969ny}. 
The Bell test experiments suffered from the locality and detection loopholes. 
The locality loophole is the ignorance of possible communication between two measurement sites. 
The detection loophole is that the non-perfect detection efficiency increases the upper bound of Bell's inequality. 
Recently, the Bell test experiments closed all loopholes, and the violation did not disappear \cite{Hensen:2015ccp}.
Therefore, the counter-intuitive prediction of Quantum Mechanics was confirmed. 
\\

\noindent
Although the violation was confirmed, its relation to Quantum Entanglement remains {\it subtle}. 
For the first step, it is necessary to improve the upper bound from a {\it quantum} generalization. 
Secondly, each entanglement quantity should diagnose the violation. 
Now a quantum generalization of the 2-qubit Bell's inequality increases the upper bound without the inconsistency \cite{Cirelson:1980ry}. 
The maximum violation is monotonically increasing with {\it concurrence} for {\it all} 2-qubit pure states as indicated by the {\it $R$-matrix} \cite{Bennett:1996gf, Wootters:1997id}. 
The {\it concurrence} is also monotonically increasing for {\it entanglement entropy}. 
It relates the degree of {\it maximum violation} to {\it Quantum Entanglement}. 
The result also establishes the {\it equivalence} between the {\it maximum violation} and the correlation of {\it $R$-matrix}.   
\\

\noindent
We hope to see the same equivalence in many-body situations. 
However, a partial trace operation only has {\it one} choice in a {\it 2-qubit} state. 
A generic {\it 2-qubit} state also only has {\it one} variable for characterizing its entanglement. 
In other words, the {\it 2-qubit} state is too {\it unusual}. 
It is {\it hard} to extend the relationship to a general $n$-qubit state \cite{Chang:2017czx, Chang:2017ygi}. 
Indeed, various difficulties of many-body Quantum Entanglement already appear in the 3-qubit state. 
Generalizing the Schmidt decomposition \cite{Peres:1994qv} shows that a general 3-qubit quantum has {\it five} independent variables \cite{Acin:2000jx}. 
Using the local operations and classical communication (LOCC) shows {\it two inequivalent} entangled classes \cite{Dur:2000zz}. 
Therefore, the degree of violation is from {\it two} entangled classes \cite{Dur:2000zz}. 
The {\it two-body} entanglement is {\it not} enough to describe the {\it tripartite} entanglement \cite{Coffman:1999jd}. 
The genuine {\it tripartite} entanglement, 3-tangle, is necessary \cite{Coffman:1999jd, Sawicki:2012}. 
The 3-qubit state shows all conceptual issues of many-body Quantum Entanglement that 2-qubit {\it cannot} answer.     
Providing the qualitative description to the 3-qubit Quantum Entanglement by Quantum Correlators should solve the universally conceptual issue of the many-body Quantum Entanglement.    
\\

\begin{figure*}
\includegraphics[width=1.\textwidth]{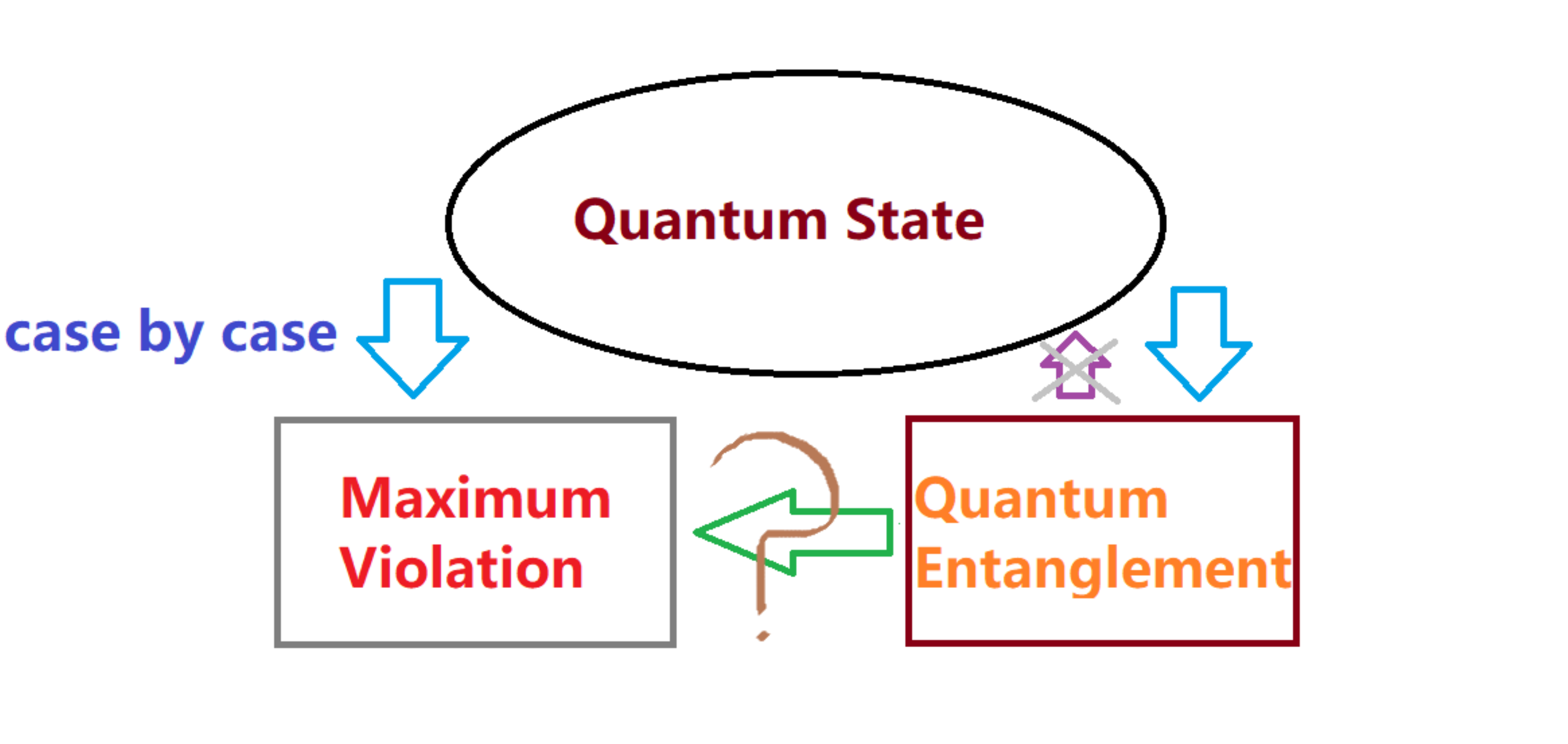}
\caption{We show the difficulties in demonstrating 3-qubit Quantum Entanglement from Mermin's inequality. 
\label{QME}}
\end{figure*}

\noindent
The central question that we would like to address in this letter is the following: {\it What is the quantitative description of the 3-qubit Quantum Entanglement through Quantum Correlator?} 
We first discuss the difficulty of building the relationship of Quantum Correlation and Entanglement from Mermin's inequality.
The $n$-qubit generalization of Bell's inequality is called {\it Mermin's inequality}. 
One can calculate the violation of Mermin's inequality {\it case by case}. 
People also know the necessary entanglement quantities. 
One can calculate {\it all} entanglement quantities for a general 3-qubit quantum state.
The problem is still there due to two difficulties as in Fig. \ref{QME}. 
The first difficulty is the {\it lack} of an analytical solution of the maximum violation of Mermin's inequality in the general 3-qubit quantum state. 
Because the state depends on five independent variables, it is {\it hard} to have an {\it inverse} relation to use entanglement quantities to express a quantum state. 
Therefore, the second difficulty is that there is possibly {\it no} analytical solution to relate the maximum violation of Mermin's inequality to entanglement quantities.  
If entanglement quantities should give a complete description of the maximum violation, it should depend on {\it five variables}. 
Similar to the first difficulty, the maximum violation should {\it not} have an analytical description for the general 3-qubit quantum state. 
Preparing a 3-qubit state is already no problem \cite{Aoki:2003, Takeda:2018}. 
Hence one inevitable task is the theoretical interpretation for a general 3-qubit state.
\\

\noindent
In this letter, we use a {\it generalized $R$-matrix} to provide an {\it analytical upper-bound} to the maximum violation of Mermin's inequality. 
We then show the rewriting of the {\it upper bound} in terms of {\it entanglement measures}. 
The fact is that Mermin's inequality loses the violation in some entangled states. 
It is not the main problem for Mermin's operator. 
For some pure states, one can determine the unique information from single-particle reduced density matrices \cite{Sawicki:2012, Maciazek:2013}. 
Here we are interested in studying the unique information of entanglement measures from some states. 
The 3-tangle case forbids the monotonically increasing result. 
Therefore, Quantum Entanglement cannot diagnose the violation of Merlin's inequality in general. 
The generalized $R$-matrix avoids the issue. 
The classification of pure many-body entanglement is one long-standing problem \cite{Sawicki:2011, Sawicki:2012(2), Sawicki:2012(3), Maciazek:2017}. 
Due to the analytical solution, we successfully {\it classify} and {\it quantify} Quantum Entanglement with the {\it experiments application}.

\section{Mermin's Inequality}
\label{sec:2}
\noindent
The Mermin's operator is 
\bea
&&
{\cal M}
\nn\\
&\equiv&
A_1\otimes A_2\otimes A_3^{\prime}
+A_1\otimes A_2^{\prime}\otimes A_3
\nn\\
&&
+A_1^{\prime}\otimes A_2\otimes A_3
-A_1^{\prime}\otimes A_2^{\prime}\otimes A_3^{\prime},
\eea
where
\bea
A_j\equiv\vec{a}_j\cdot \vec{\sigma}; \qquad A_j^{\prime}\equiv\vec{a}_j^{\prime}\cdot\vec{\sigma}; \qquad 
\vec{\sigma}\equiv(\sigma_x, \sigma_y, \sigma_z).
\eea 
The $\vec{a}$ and $\vec{a}^{\prime}$ are the unit vectors.
For any 3-qubit state, the upper bound of the expectation value of the Mermin's operator is:
\bea
\langle{\cal M}\rangle\equiv\mathrm{Tr}(\rho{\cal M})\le 4, 
\eea
where $\rho\equiv|\psi\rangle\langle\psi|$ is a density matrix. 
Mermin's inequality is $\langle {\cal M}\rangle\le 2$. 
Therefore, the quantum state shows the violation when the $\langle {\cal M}\rangle$ is larger than 2. 
The maximum violation monotonically increases for the concurrence in 2-qubit. 
The different choice of Mermin's operator should provide a different quantification to Quantum Entanglement. 
Therefore, considering all possible choices of Mermin's operator 
\bea
\gamma\equiv\max_{{\cal M}}\langle{\cal M}\rangle
\eea
 should be proper to demonstrate Quantum Entanglement because it is independent of a partial trace operation. 
\\

\noindent
The expectation value of the Mermin's operator is given by: 
\bea
&&
\langle {\cal M}\rangle
\nn\\
&=&
\bigg(a_1, a_2^TRa_3^{\prime}\bigg)
+\bigg(a_1, a_2^{\prime T}Ra_3\bigg)
\nn\\
&&
+\bigg(a_1^{\prime}, a_2^TRa_3\bigg)
-\bigg(a_1^{\prime}, a_2^{\prime T}Ra_3^{\prime}\bigg),
\eea
where $a_j\equiv (a_{j, x}, a_{j, y}, a_{j, z})^T$. 
The superscript $T$ refers to the transpose. 
The definition of $a_j^{\prime}$ is similar to the $a_j$.
The $R\equiv (R_x, R_y, R_z)$ is the generalized $R$-matrix. 
Each element of the generalized $R$-matrix is defined as $R_j\equiv (R_{jkm})$.  
The inner product is defined as 
\bea
\big(a_1, a_2^TRa_3^{\prime}\big)
\equiv
\sum_{i_1, i_2, i_3=1}^3 a_{1, i_1}a_{2, i_2}a_{3, i_3}^{\prime}R_{i_1 i_2 i_3}, 
\eea
where $R_{j_1j_2j_3}\equiv\mathrm{Tr}(\rho\sigma_{j_1}\otimes\sigma_{j_2}\otimes\sigma_{j_3})$.
Note that the following vectors are orthogonal:
\bea
V&\equiv& \begin{pmatrix}
a_{2, j}a_{3, k}^{\prime}+a_{2, j}^{\prime}a_{3, k}
\end{pmatrix}; 
\nn\\
V^{\prime}&\equiv&
\begin{pmatrix}
a_{2, j}a_{3, k}-a_{2, j}^{\prime}a_{3, k}^{\prime}
\end{pmatrix}.
\eea
The norm of two vectors is:
\bea
|V|^2&=&2+2\cos(\theta_2)\cos(\theta_3); 
\nn\\
 |V^{\prime}|^2&=&2-2\cos(\theta_2)\cos(\theta_3),
\eea
where $\vec{a}_l\cdot\vec{a^{\prime}}_l\equiv\cos(\theta_l)$. 
The range of $\theta_l$ is $0\le\theta_l\le\pi$. 
Here we define $\cos(2\theta)\equiv\cos(\theta_2)\cos(\theta_3)$, where $0\le\theta\le\pi/2$.
Therefore, we introduce the orthogonal unit-vectors, $c$; $c^{\prime}$, as in the following:
\bea
V\equiv 2c\cos(\theta); \qquad V^{\prime}\equiv 2c^{\prime}\sin(\theta).
\eea
Therefore, we rewrite the formula as $\langle {\cal M}\rangle=
2\cos(\theta)\big( a_1, R c\big)+2 \sin(\theta)\big(a_1^{\prime}, R c^{\prime}\big)$. 
The matrix multiplication of $RR^T$ has three possible but not equivalent choices in general:
\bea
R^{(1)}_{j_1J_1}&\equiv& R_{j_1j_2j_3}|_{J_1=(j_2, j_3)}; 
\nn\\
 R^{(2)}_{j_2J_2}&\equiv& R_{j_1j_2j_3}|_{J_2=(j_1, j_3)}; 
 \nn\\
  R^{(3)}_{j_3J_3}&\equiv& R_{j_1j_2j_3}|_{J_3=(j_1, j_2)}, 
\eea
where $j_1, j_2, j_3=x, y, z$.
The different matrix multiplications give different results. 
Hence we obtain: 
\bea
\gamma\le\gamma_R=2\min_{R^{(1)}, R^{(2)}, R^{(3)}}\sqrt{u_1^2+u_2^2},
\eea
 where the $u_1^2$ and $u_2^2$ are the two largest eigenvalues of $RR^T$. 
Here we used the following inequality $|x_1^T{\cal B}x_2|\le \lambda|x_1||x_2|$. 
The $x_1$ is the $m$-dimensional vector, and $x_2$ is the $n$-dimensional vector. 
The ${\cal B}$ is an $m\times n$ rectangular matrix, and $\lambda$ is the maximum singular value of ${\cal B}$.
The inequality saturates the upper bound only when the vectors, $x_1$ and $x_2$, are the corresponding singular vectors for the $\lambda$. 
We will show that $\gamma_R$ is not equivalent to $\gamma$ in general. 

\section{Quantum Entanglement and Quantum Correlation}
\label{sec:3}
\noindent
A general 3-qubit quantum state up to a local unitary transformation is \cite{Acin:2000jx} 
\bea
&&
|\psi\rangle
\nn\\
&=&\lambda_0|000\rangle
+\lambda_1e^{i\phi}|100\rangle
+\lambda_2|101\rangle
+\lambda_3|110\rangle
+\lambda_4|111\rangle.
\nn\\
\eea
The $\lambda_j$ is non-negative. 
The range of $\phi$ is $0\le\phi\le\pi$. 
The normalization of a density matrix $\mathrm{Tr}\rho=1$ provides the normalization $\lambda_0^2+\lambda_1^2+\lambda_2^2+\lambda_3^2+\lambda_4^2=1$. 
Hence a general 3-qubit quantum state only has five independent variables. 
\\

\noindent
Using the 3-qubit quantum state shows that the eigenvalues of $RR^T\equiv M$ follows from the equation $x^3+\alpha_1^{(j)}x^2+\alpha_2^{(j)}x+\alpha_3^{(j)}=0$. 
The solution $x$ is the eigenvalue of $RR^T$. 
The superscript $(j)$ indicates the choice of $RR^T$ multiplication. 
The $\alpha_1^{(j)}, \alpha_2^{(j)}, \alpha_3^{(j)}$ are real-valued. 
The eigenvalues are also real-valued.  
Hence the discriminant is not a positive number: 
\bea
&&
\Delta^{(j)}
\nn\\
&\equiv&\bigg(-\frac{\big(\alpha_1^{(j)}\big)^{3}}{27}-\frac{\alpha_3^{(j)}}{2}+\frac{\alpha_1^{(j)}\alpha_2^{(j)}}{6}\bigg)^2
\nn\\
&&
+\bigg(\frac{\alpha_2^{(j)}}{3}-\frac{\big(\alpha_1^{(j)}\big)^2}{9}\bigg)^3
\nn\\
&\equiv&\big(\gamma_1^{(j)}\big)^2+\big(\gamma_2^{(j)}\big)^3\le 0.
\eea
The eigenvalues are: 
\bea
&&
x_1^{(j)}
\nn\\
&=&-\frac{\alpha_1^{(j)}}{3}+2\sqrt{-\gamma^{(j)}_2}\cos\bigg\lbrack
\frac{1}{3}\arccos\bigg(\frac{\gamma_1^{(j)}}{(-\gamma_2^{(j)})^{\frac{3}{2}}}\bigg)\bigg\rbrack; 
\nn\\
&&
x_2^{(j)}
\nn\\
&=&-\frac{\alpha_1^{(j)}}{3}+2\sqrt{-\gamma^{(j)}_2}\cos\bigg\lbrack
\frac{1}{3}\arccos\bigg(\frac{\gamma_1^{(j)}}{(-\gamma_2^{(j)})^{\frac{3}{2}}}\bigg)+\frac{2\pi}{3}\bigg\rbrack; 
\nn\\
&&
x_3^{(j)}
\nn\\
&=&-\frac{\alpha_1^{(j)}}{3}+2\sqrt{-\gamma^{(j)}_2}\cos\bigg\lbrack
\frac{1}{3}\arccos\bigg(\frac{\gamma_1^{(j)}}{(-\gamma_2^{(j)})^{\frac{3}{2}}}\bigg)-\frac{2\pi}{3}\bigg\rbrack.
\nn\\
\eea
Therefore, the variables, $\alpha_1^{(j)}; \alpha_2^{(j)}; \alpha_3^{(j)}$, fully determines $\gamma_R$.
In other words, we successfully establish the relation of Quantum Correlation and Quantum Entanglement.
\\

\noindent
For a 3-qubit quantum state, all invariant quantities are the following:
\bea
&&
I_1=\mathrm{Tr}\rho_1^2; \qquad 
I_2=\mathrm{Tr}\rho_2^2; \qquad 
I_3=\mathrm{Tr}\rho_3^2; 
\nn\\
&&
I_4=\tau_{1|23}-\tau_{1|2}-\tau_{1|3}; 
\nn\\
&&
I_5=\mathrm{Tr}\big((\rho_1\otimes\rho_2)\rho_{12}\big)-\frac{1}{3}\mathrm{Tr}(\rho_1^3)-\frac{1}{3}\mathrm{Tr}(\rho_2^3),
\eea
where $\rho_j$ is the reduced density matrix of the $j$-th qubit. 
The $\tau_{1|23}\equiv 2(1-\mathrm{Tr}\rho_1^2)$ \cite{Wootters:1997id}. 
The $\sqrt{\tau_{i_1|i_2}}$ is the entanglement of formation of the $i_1$ qubit and $i_2$ qubit \cite{Wootters:1997id}. 
The entanglement of formation is defined as the following \cite{Bennett:1996gf, Wootters:1997id}:
\bea
C(\rho)&\equiv&\min_{p_j, \psi_j}\sum_j p_jC(\psi_j)=\max(0, Q_1-Q_2-Q_3-Q_4), 
\nn\\
&&
Q_1\ge Q_2\ge Q_3\ge Q_4; 
\nn\\
\rho&=&\sum_jp_j|\psi_j\rangle\langle\psi_j|,
\eea   
where $Q_j$ is the eigenvalue of $\sqrt{\rho(\sigma_y\otimes\sigma_y)\rho^*(\sigma_y\otimes\sigma_y)}$ \cite{Bennett:1996gf, Wootters:1997id}. The $*$ is the complex conjugate.
The minimization is the overall decompositions of the density matrix $\rho$.
The $I_4$ is called 3-tangle \cite{Coffman:1999jd}. 
The $I_5$ is a combination of the correlation of the reduced density matrix of the first qubit and second qubit $\mathrm{Tr}\big((\rho_1\otimes\rho_2)\rho_{12}\big)-\mathrm{Tr}(\rho_1^2)-\mathrm{Tr}(\rho_2^2)$ and other invariant quantities. 
We choose the following entanglement quantities: 
\bea
&&
E_1\equiv\tau_{1|2}; \qquad 
E_2\equiv\tau_{1|3}; \qquad 
E_3\equiv\tau_{2|3}; \qquad 
\nn\\
&&
E_4\equiv\tau=I_4; \qquad 
E_5\equiv I_5+\frac{1}{4}(E_1^2+E_2^2+E_4^2)+E_3^2
\nn\\
\eea
because the $E_5$ is invariant for the different $RR^T$-multiplication, but the $I_5$ does not. 
\\

\noindent
The $E_1, E_2, E_3$ are the functions of the concurrences $C_1, C_2, C_3$. 
The $C_j\equiv\sqrt{2(1-\mathrm{Tr}\rho_j^2)}$ is the concurrence of the $j$-th qubit. 
By the relation, using the $E_1, E_2, \cdots, E_5$ is the same as using the $I_1, I_2, \cdots, I_5$.  
Now we use the $R^{(1)}$ to calculate the $\alpha_1^{(1)}, \alpha_2^{(1)}, \alpha_3^{(1)}$:
\bea
&&
\alpha_1^{(1)}
\nn\\
&=&-1-(2E_1^2+2E_2^2+2E_3^2+3E_4^2)
\nn\\
&=&-1-(C_1^2+C_2^2+C_3^2)
\nn\\
&\equiv&-1-C_T^2; 
\nn\\
&&
\alpha_2^{(1)}
\nn\\
&=&2(E_1^2+E_2^2+E_4^2)E_3^2+2(E_1^2+E_2^2)(E_4^2+1)
\nn\\
&&
+E_1^4+E_2^4+4E_4^2+16E_5;
\nn\\
&&
\alpha_3^{(1)}
\nn\\
&=&(E_1^2+E_2^2+2E_3^2+2E_4^2)
\nn\\
&\times&(2E_4^4+2E_1^2E_2^2+E_1^2E_4^2+E_2^2E_4^2)
\nn\\
&&
-(E_1^2+E_2^2+2E_4^2+8E_5)^2.
\eea
The $C_T$ is called total concurrence. 
This entanglement measure is also invariant for different multiplication ways of $RR^T$. 
We exchange $E_2$ and $E_3$ to obtain $\alpha_2^{(2)}$ and $\alpha_3^{(2)}$. 
For exchanging $E_1$ and $E_3$, we get $\alpha_2^{(3)}$ and $\alpha_3^{(3)}$.
It is easy to show that $\alpha_1^{(j)}$ is negative, and $\alpha_2^{(j)}$ is non-negative. 
Hence the upper bound $\gamma_R$ contains all necessary entanglement measures.
\\

\noindent 
Now we show the analytical solution to the maximum violation of Mermin's inequality. 
Due to the inequality: 
\bea
0\le\theta^{(j)}\equiv\frac{1}{3}\arccos\bigg(\frac{\gamma^{(j)}_1}{(-\gamma^{(j)}_2)^{\frac{3}{2}}}\bigg)\le\frac{\pi}{3},
\eea
 only the first eigenvalue $x^{(j)}_1$ is not negative for all possible variables. 
The inequality also implies $x^{(j)}_3\ge x^{(j)}_2$. 
Therefore, the calculation of $\gamma_R$ always chooses $x_1^{(j)}$ and $x_3^{(j)}$ for a general 3-qubit quantum state. 
Hence we obtain 
\bea
\gamma_R=2\min_j\sqrt{-\frac{2\alpha_1^{(j)}}{3}+2\sqrt{-\gamma^{(j)}_2}\cos\bigg(\theta^{(j)}-\frac{\pi}{3}\bigg)}.
\nn\\
\eea 
Indeed, it does not have a global monotonically increasing function for $\gamma_R$. 
It is consistent with the LOCC \cite{Dur:2000zz}. 
The 3-qubit quantum state has two inequivalent entangled classes, GHZ- and W-state \cite{Dur:2000zz}. 
In other words, quantifying 3-qubit Quantum Entanglement should at least need three parameters. 
Two parameters are for the overlapping level between a state and the GHZ-state and W-state. 
The remaining one is for quantifying Quantum Entanglement. 
Hence we should fix two parameters for the quantification and consistency of LOCC.
We find that $\gamma_R$ is monotonically increasing for $-\alpha_1^{(j)}$ without varying $\gamma_2^{(j)}$ and $\theta^{(j)}$. 
The $\alpha_1^{(j)}$ is invariant for the different $RR^T$-multiplication. 
It should be suitable to quantify Quantum Entanglement. 
Hence we can use $\gamma_2^{(j)}$ and $\theta^{(j)}$ to classify Quantum Entanglement. 
We will show one figure to demonstrate when we compare the $\gamma_R$ to $\gamma$. 
\\

\begin{figure}[!htb]
\begin{center}
\includegraphics[width=1.\textwidth]{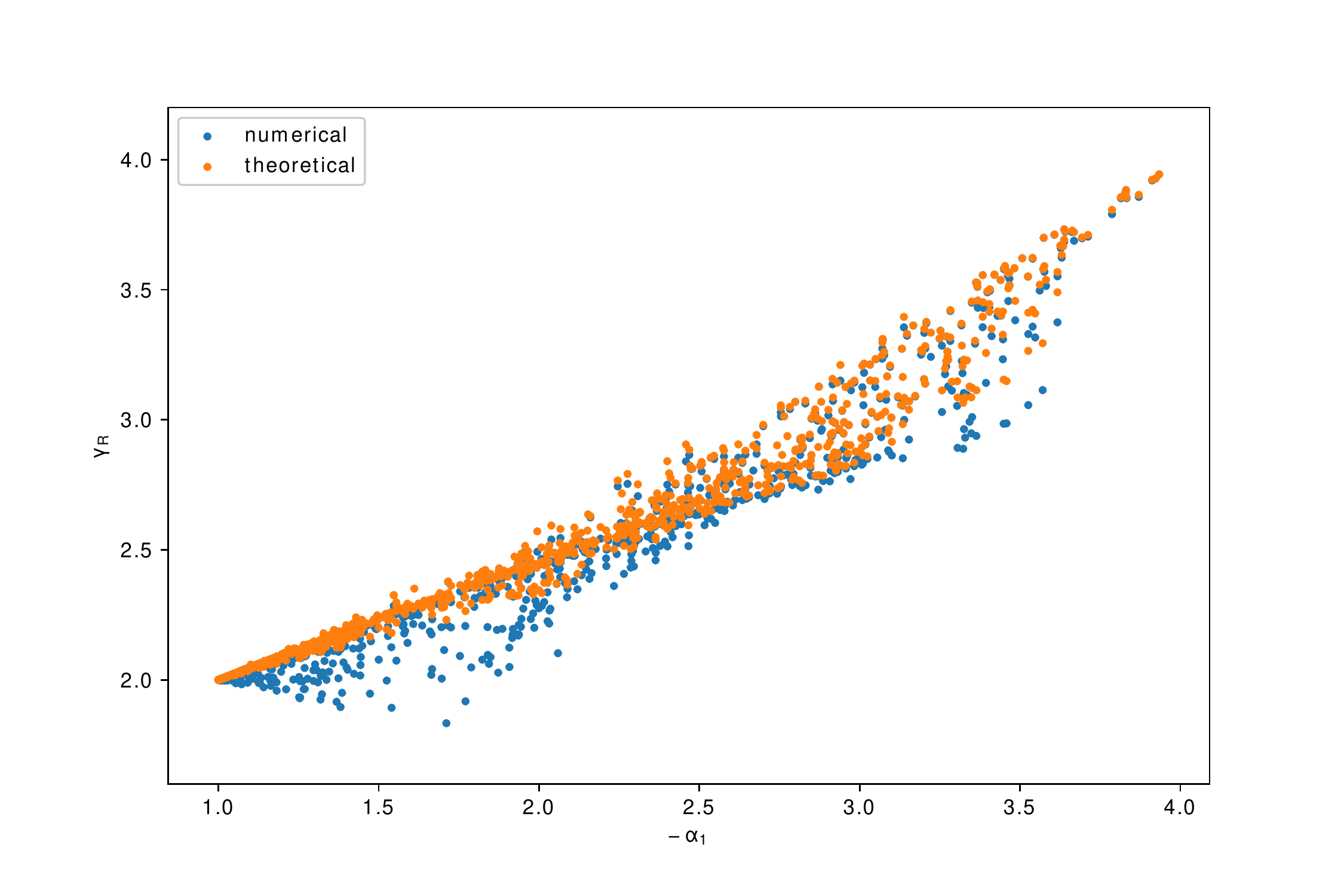}
\caption{Theoretical upper bound (orange) and numerical results (blue) versus $-\alpha_1$ ($x$-axis).}\label{Fig:general}
\end{center}
\end{figure}

\begin{figure}[!htb]
\begin{center}
\includegraphics[width=1.\textwidth]{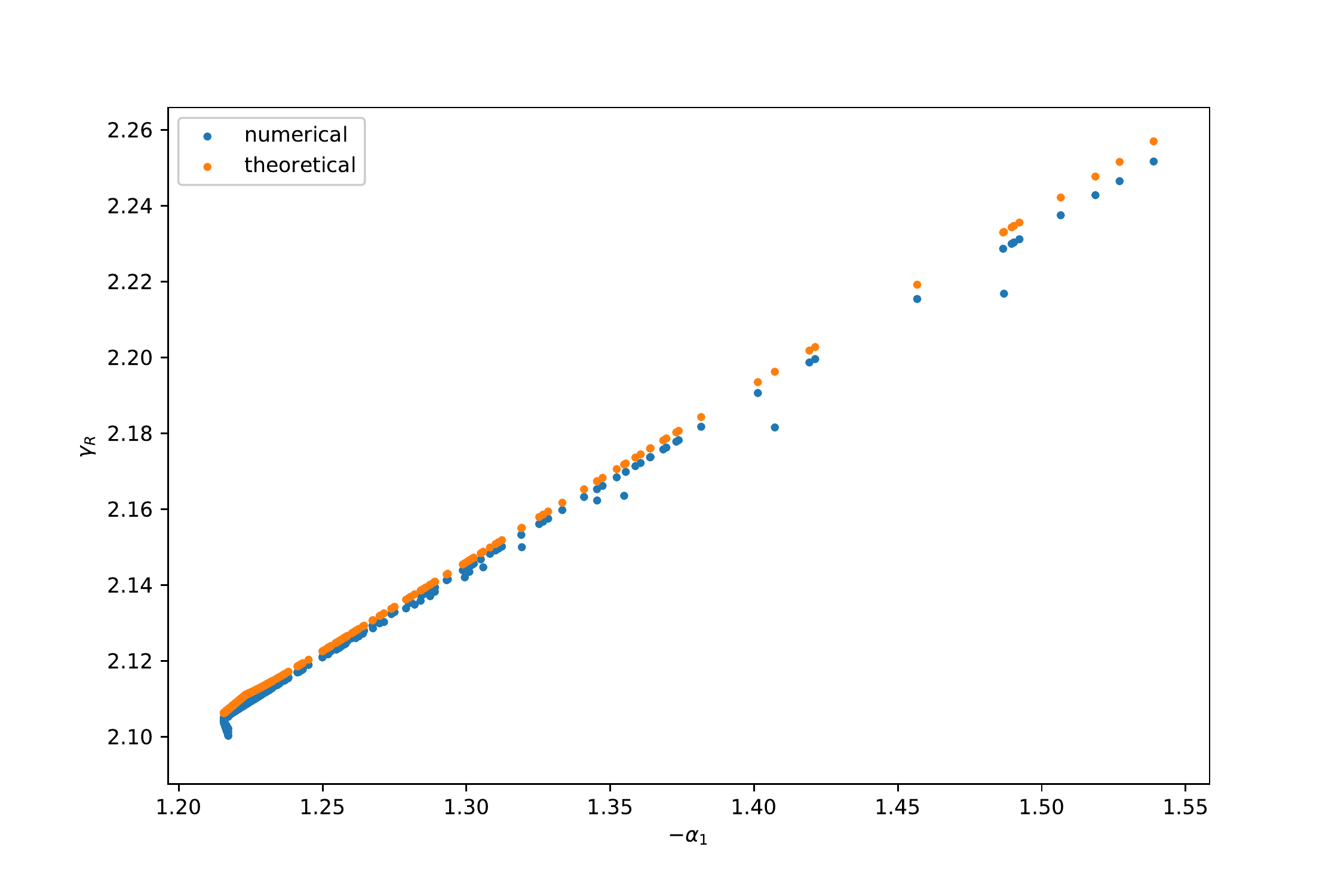}
\caption{Theoretical upper bound (orange) and numerical results (blue) versus $-\alpha_1$ ($x$-axis) with the fixed parameters $\gamma_2^{(j)}=-0.0884$ and $\theta^{(j)}=0.00238$.} 
\label{Fig:fix}
\end{center}
\end{figure}

\noindent
We perform the optimization on $\langle {\cal M}\rangle$ using a direct numerical calculation. 
We then prepare $1000$ 3-qubit quantum states. 
Fig. \ref{Fig:general} shows the differences between the $\gamma$ and $\gamma_R$. 
Because $-\alpha_1^{(j)}$ does not depend on the index $j$, we remove the index $j$ in this figure. 
We also choose $\lambda_4=0$ to demonstrate the classification in Fig. \ref{Fig:fix}. 
Here we can do a more detailed analysis but less trivial than the two-body case (the $\gamma_R$ only depends on $E_1$, $E_2$, and $E_3$). 
We also compare the $\gamma_R$ to $\gamma$ in this figure. 
The $\gamma_R$ is monotonically increasing for $-\alpha_1$ as expected, but the $\gamma$ does not.
It shows the difference again. 
Because the classification is relevant to complicated functions, it is hard to find. 
Hence we demonstrate the usefulness of the analytical solution. 
\\

\noindent
In the two-body case, the correlation of the $R$-matrix provides the maximum violation. 
We generalize the $R$-matrix from Mermin's inequality and relate the matrix to entanglement measures as in the two-qubit case. 
The generalized $R$-matrix cannot show the maximum violation in general. 
Our result provides a concrete realization for relating Quantum Correlation to Quantum Entanglement. 
Two-Body Quantum Entanglement only depends on one independent variable. 
One can rewrite the maximum violation in terms of concurrence. 
The rewriting can also be inverted. 
Because a general three-qubit quantum state has five independent variables, we lose the inverse relation. 
Measuring entanglement measures by experiment correlators becomes difficult. 
We will reduce a generic 3-qubit problem to one entanglement measure. 
It is helpful for the experimental realization of our theoretical study. 
In Fig. \ref{Fig:fix}, we find no monotonically increase in $\gamma$. 
It suggests that Quantum Entanglement cannot induce the violation of Mermin's inequality. 
Later we will move to one entanglement measure. 
We will show no monotonic increasing behavior.
It shows the impossibility of demonstrating Quantum Entanglement by the degree of violation of Mermin's inequality.

\section{Experiment Correlator and Tripartite Entanglement}
\label{sec:4}
\begin{figure}[!htb]
\begin{center}
\includegraphics[width=1.\textwidth]{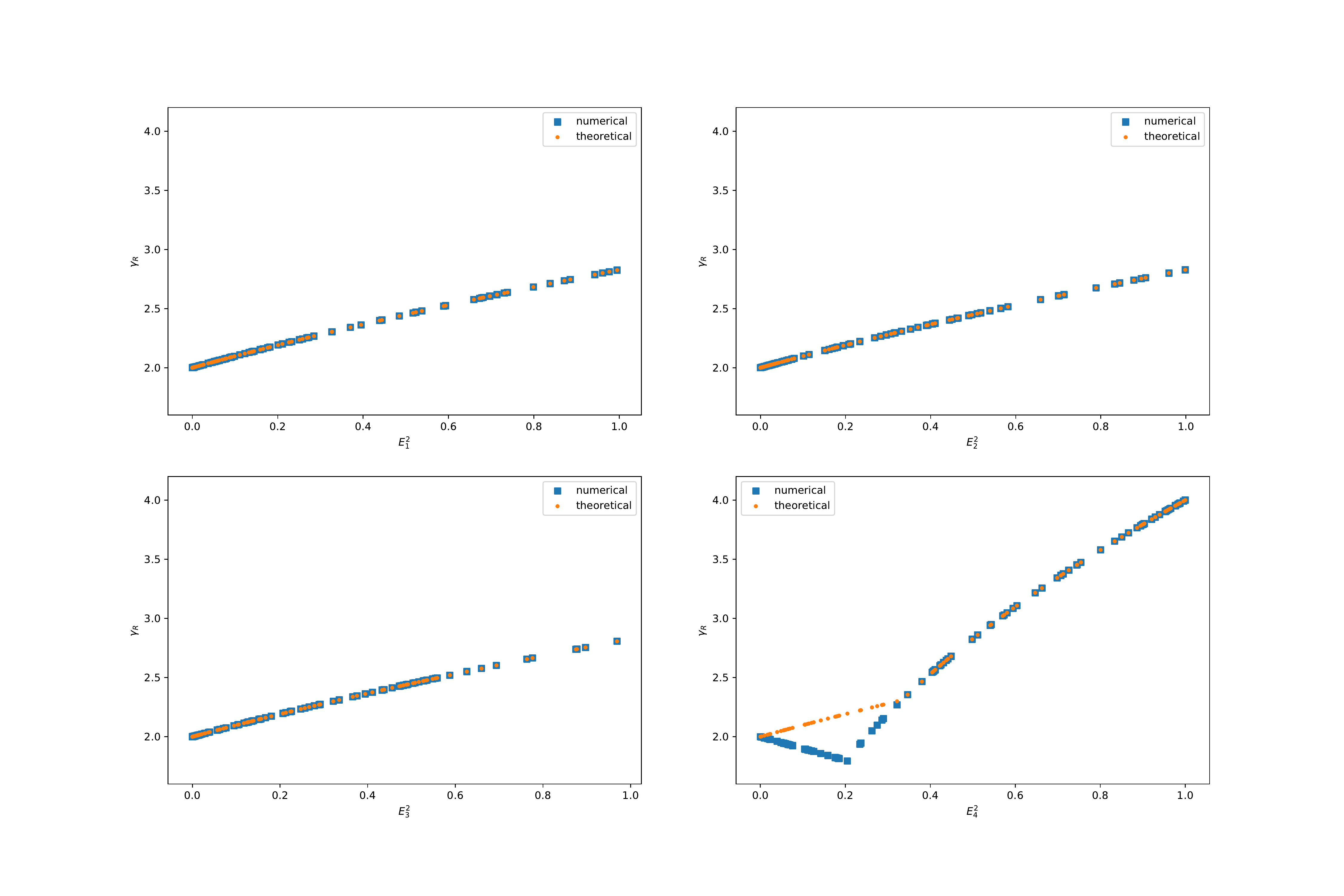}
\caption{Theoretical (orange) and numerical results (blue) versus $E_1^2$ (upper left), $E_2^2$ (upper right), $E_3^2$ (lower left), and $E_4^2$ (lower right).}
\label{Fig:oneE}
\end{center}
\end{figure}

\noindent 
We turn off $\lambda_2$ and $\lambda_4$
then only leaves the non-vanishing $E_1$.  
When we consider $\lambda_3=\lambda_4=0$,
the only non-vanishing entanglement quantity is $E_2$. 
For the $E_3$, we turn off the $\lambda_0$. 
Finally, we choose the $\lambda_1=\lambda_2=\lambda_3=0$ to measure 3-tangle. 
Fig. \ref{Fig:oneE} shows the monotonically increasing behavior of $\gamma_R$ for the $E_1, E_2, E_3, E_4$. 
We find that the results are identical for $E_1, E_2, E_3$, while the result for $E_4$ is different. 
The mismatch at the small $E_4$ region corresponds to the small $-\alpha_1$ region in Fig. \ref{Fig:general}. 
With only $\lambda_0$ and $\lambda_4$, the 3-qubit state is the GHZ class, $\lambda_0|000\rangle+\lambda_4|111\rangle$.  
The GHZ class is different from the other 3 cases (at least one qubit is the same). 
We can check the first three cases. 
The $\theta^{(j)}$ remains 0.  
The $\gamma_R=\gamma$ is monotonically increasing with entanglement quantity as in $2\sqrt{1+E_j^2}$. 
While in the GHZ class, the $\theta^{(j)}$ jumps from 0 to $\pi/3$ at $E_4^2=1/3$ in the theoretical upper bound. 
Hence it shows the difference of the $\gamma_R$ and $\gamma$ from the non-vanishing $\theta^{(j)}$. 
The numerical solution shows that some entangled states do not have the violation. 
However, some entangled states also have smaller values than the product state in the $\gamma_R$. 
Hence no violation of inequality in an entangled state is not the main issue for Mermin's operator. 
The real problem occurs when a state only depends on one entanglement measure. 
For this case, the monotonically increasing behavior cannot lose.
Otherwise, it is impossible to quantify Quantum Entanglement. 
Hence Quantum Entanglement cannot be a source of violation for Mermin's inequality. 
Choosing other operators can show the monotonically increasing behavior in the GHZ class \cite{Chang:2017czx, Chang:2017ygi}. 
However, the proposal is also not general. 
Because the three-qubit quantum state has two inequivalent entangled classes, the result possibly implies no generic operators to show the violation from Quantum Entanglement.  
\\

\noindent
Because we already have complete information for the $E_1, E_2, E_3, E_4$ from the one entanglement measure case, the only unknown quantity is $E_5$. 
One can measure various quantum states to extract $E_5$ from $\alpha_2^{(1), (2), (3)}$ and $\alpha_3^{(1), (2), (3)}$. 
Because $E_5$ is relevant to the correlation between the reduced density matrices, it should be interesting. 

\section{Outlook}
\label{sec:5}
\noindent
One difficulty of many-body Quantum Entanglement is too many independent variables. 
A 3-qubit quantum state has five independent variables. 
Naively, One should expect a quintic equation for relating the correlators to Quantum Entanglement. 
Our theoretical study showed that naive expectation is wrong. 
Because $RR^T$ is a three by three matrix, the cubic equation is enough. 
Therefore, we could show an analytical solution to the $\gamma_R$.
The five necessary entanglement variables all appear in the $\gamma_R$. 
\\

\noindent
It has a general expectation that the violation is a diagnosis of quantumness. 
Our 3-qubit study should suggest ``Violation$\neq$ Quantum''. 
We proposed an alternative diagnosis, the generalized $R$-matrix.   
Developing a generic diagnosis to an $n$-qubit state should be a revolutionary breakthrough. 
\\

\noindent
We provided a classification to quantify Quantum Entanglement by Quantum Correlators. 
A partial trace operation is unnecessary for measuring $\gamma_R$. 
A partial trace operation leads to a hard-measuring problem to entanglement quantities.
Therefore, measuring $\gamma_R$ is possible. 
Because the correlator is measurable \cite{Aoki:2003}, the classification is realizable. 
Therefore, our study provided an alternative measure by the correlators.  

\section*{Acknowledgments}
\noindent 
We thank Xing Huang, Ling-Yan Hung, Masaki Tezuka, and Shanchao Zhang for their helpful discussion. 
Chen-Te Ma would like to thank Nan-Peng Ma for his encouragement.
\\

\noindent
Xingyu Guo acknowledges the Guangdong Major Project of Basic and Applied Basic Research No. 2020B0301030008 and NSFC Grant No.11905066.
Chen-Te Ma acknowledges the China Postdoctoral Science Foundation, Postdoctoral General Funding: Second Class (Grant No. 2019M652926); 
Foreign Young Talents Program (Grant No. QN20200230017); 
Post-Doctoral International Exchange Program.



\baselineskip 22pt
\begin{thebibliography}{99}

\bibitem{Bell:1964kc}
J.~S.~Bell,
``On the Einstein-Podolsky-Rosen paradox,''
Physics Physique Fizika \textbf{1}, 195-200 (1964)
doi:10.1103/PhysicsPhysiqueFizika.1.195

\bibitem{Clauser:1969ny}
J.~F.~Clauser, M.~A.~Horne, A.~Shimony and R.~A.~Holt,
``Proposed experiment to test local hidden variable theories,''
Phys. Rev. Lett. \textbf{23}, 880-884 (1969)
doi:10.1103/PhysRevLett.23.880

\bibitem{Hensen:2015ccp}
B.~Hensen, H.~Bernien, A.~E.~Dreau, A.~Reiserer, N.~Kalb, M.~S.~Blok, J.~Ruitenberg, R.~F.~L.~Vermeulen, R.~N.~Schouten and C.~Abellan, \textit{et al.}
``Loophole-free Bell inequality violation using electron spins separated by 1.3 kilometres,''
Nature \textbf{526}, 682-686 (2015)
doi:10.1038/nature15759
[arXiv:1508.05949 [quant-ph]].

\bibitem{Cirelson:1980ry}
B.~S.~Cirelson,
``QUANTUM GENERALIZATIONS OF BELL'S INEQUALITY,''
Lett. Math. Phys. \textbf{4}, 93-100 (1980)
doi:10.1007/BF00417500

\bibitem{Bennett:1996gf}
C.~H.~Bennett, D.~P.~DiVincenzo, J.~A.~Smolin and W.~K.~Wootters,
``Mixed state entanglement and quantum error correction,''
Phys. Rev. A \textbf{54}, 3824-3851 (1996)
doi:10.1103/PhysRevA.54.3824
[arXiv:quant-ph/9604024 [quant-ph]].

\bibitem{Wootters:1997id}
W.~K.~Wootters,
``Entanglement of formation of an arbitrary state of two qubits,''
Phys. Rev. Lett. \textbf{80}, 2245-2248 (1998)
doi:10.1103/PhysRevLett.80.2245
[arXiv:quant-ph/9709029 [quant-ph]].

\bibitem{Chang:2017czx}
P.~Y.~Chang, S.~K.~Chu and C.~T.~Ma,
``Bell's Inequality and Entanglement in Qubits,''
JHEP \textbf{09}, 100 (2017)
doi:10.1007/JHEP09(2017)100
[arXiv:1705.06444 [quant-ph]].

\bibitem{Chang:2017ygi}
P.~Y.~Chang, S.~K.~Chu and C.~T.~Ma,
``Bell\textquoteright{}s inequality, generalized concurrence and entanglement in qubits,''
Int. J. Mod. Phys. A \textbf{34}, no.06n07, 1950032 (2019)
doi:10.1142/S0217751X19500325
[arXiv:1710.10493 [quant-ph]].

\bibitem{Peres:1994qv}
A.~Peres,
``Higher order Schmidt decompositions,''
Phys. Lett. A \textbf{202}, 16-17 (1995)
doi:10.1016/0375-9601(95)00315-T
[arXiv:quant-ph/9504006 [quant-ph]].

\bibitem{Acin:2000jx}
A.~Acin, A.~A.~Andrianov, L.~Costa, E.~Jane, J.~I.~Latorre and R.~Tarrach,
``Generalized Schmidt Decomposition and Classification of Three-Quantum-Bit States,''
Phys. Rev. Lett. \textbf{85}, 1560-1563 (2000)
doi:10.1103/PhysRevLett.85.1560
[arXiv:quant-ph/0003050 [quant-ph]].

\bibitem{Dur:2000zz}
W.~Dur, G.~Vidal and J.~I.~Cirac,
``Three qubits can be entangled in two inequivalent ways,''
Phys. Rev. A \textbf{62}, 062314 (2000)
doi:10.1103/PhysRevA.62.062314
[arXiv:quant-ph/0005115 [quant-ph]].

\bibitem{Coffman:1999jd}
V.~Coffman, J.~Kundu and W.~K.~Wootters,
``Distributed entanglement,''
Phys. Rev. A \textbf{61}, 052306 (2000)
doi:10.1103/PhysRevA.61.052306
[arXiv:quant-ph/9907047 [quant-ph]].

\bibitem{Sawicki:2012}
A.~Sawicki, M.~Walter and M.~Kuś,
``When is a pure state of three qubits determined by its single-particle reduced density matrices?,''
J. Phys. A \textbf{46}, 055304 (2013)
doi:10.1088/1751-8113/46/5/055304
[arXiv:1207.3849 [quant-ph]]. 

\bibitem{Aoki:2003}
T.~Aoki, N.~Takei, H.~Yonezawa, K.~Wakui, T.~Hiraoka, A.~Furusawa and P.~van Loock,
``Experimental Creation of a Fully Inseparable Tripartite Continuous-Variable State,''
Phys. Rev. Lett. \textbf{91}, 080404 (2003)
doi:10.1103/PhysRevLett.91.080404
[arXiv:quant-ph/0304053 [quant-ph]]. 

\bibitem{Takeda:2018}
S.~Takeda, K.~Takase and A.~Furusawa,
``On-demand photonic entanglement synthesizer,''
Science Advances \textbf{5}, eaaw4530 (2019)
doi:10.1126/sciadv.aaw4530
[arXiv:1811.10704 [quant-ph]]. 

\bibitem{Maciazek:2013}
T.~Maciazek, M.~Oszmaniec and A.~Sawicki,
``How many invariant polynomials are needed to decide local unitary equivalence of qubit states?,''
J. Math. Phys. \textbf{54}, 092201 (2013) 
doi:10.1063/1.4819499
[arXiv:1305.3894 [quant-ph]].

\bibitem{Sawicki:2011}
A.~Sawicki and M.~Kuś,
``Geometry of the local equivalence of states,''
J. Phys. A \textbf{44}, 495301 (2011) 
doi:10.1088/1751-8113/44/49/495301
[arXiv:1108.4134 [math-ph]]. 

\bibitem{Sawicki:2012(2)}
A.~Sawicki, M.~Oszmaniec and M.~Kuś,
``Convexity of momentum map, Morse index, and quantum entanglement,''
Reviews in Mathematical Physics \textbf{26}, 1450004 (2014) 
doi:10.1142/S0129055X14500044
[arXiv:1208.0556 [math-ph]]. 

\bibitem{Sawicki:2012(3)}
A.~Sawicki, M.~Oszmaniec and M.~Kuś,
``Critical sets of the total variance of state detect all SLOCC entanglement classes,''
Phys. Rev. A \textbf{86}, 040304(R) (2012) 
doi:10.1103/PhysRevA.86.040304
[arXiv:1208.0557 [math-ph]]. 

\bibitem{Maciazek:2017}
T.~Maciażek and A.~Sawicki,
``Asymptotic properties of entanglement polytopes for large number of qubits,''
J. Phys. A \textbf{51}, 07LT01 (2018) 
doi:10.1088/1751-8121/aaa4d7
[arXiv:1706.05019 [quant-ph]]. 

\end{thebibliography}
\end{document}